\documentclass[%
 aip,
 amsmath,amssymb,
 reprint,%
]{revtex4-1}

\usepackage{graphicx}% Include figure files
\usepackage{dcolumn}% Align table columns on decimal point
\usepackage{bm}% bold math
\usepackage[utf8]{inputenc}
\usepackage[T1]{fontenc}
\usepackage{mathptmx}
\usepackage{etoolbox}

\makeatletter
\def\@email#1#2{%
 \endgroup
 \patchcmd{\titleblock@produce}
  {\frontmatter@RRAPformat}
  {\frontmatter@RRAPformat{\produce@RRAP{*#1\href{mailto:#2}{#2}}}\frontmatter@RRAPformat}
  {}{}
}%
\makeatother

\newcommand{\Tr}{\operatorname{Tr}}

\newtheorem{mydef}{Definition}

\usepackage{hyperref}

\begin{document}

\preprint{AIP/123-QED}

\title[Application of $\alpha$-order Information Metrics]{Application of $\alpha$-order Information Metrics for Secure Communication in Quantum Physical Layer Design}
% Force line breaks with \\
\author{Masahito Hayashi}
 \affiliation{The Chinese University of Hong Kong, Shenzhen, Longgang District, Shenzhen, 518172, China}%Lines break automatically or can be forced with \\
 \altaffiliation[Also at ]{International Quantum Academy, Futian District, Shenzhen 518048, China,
and Graduate School of Mathematics, Nagoya University, Japan.}%Lines break automatically or can be forced with \\
\author{Angeles Vázquez-Castro}%
\affiliation{Dpt. of Telecommunications and Systems Engineering, 
  School of Engineering, 
  Autonomous University of Barcelona, 
08193 Bellaterra Campus, Barcelona, Spain}%
\altaffiliation[Also at ]{Institute of Space Studies of Catalonia (IEEC-UAB)
Mediterranean Technology Park, 
08860 Castelldefels, Barcelona, Spain}
 \email{hmasahito@cuhk.edu.cn}
 \email{angeles.vazquez@uab.cat}
\date{\today}% It is always \today, today,

\begin{abstract}
Quantum physical layer security protocols offer significant advantages, particularly in space communications, but their achievable secrecy rates are often limited to the asymptotic regime. Realistic communication systems, however, demand non-asymptotic solutions. Recently, the $\alpha$-order information-theoretic metrics based on Rényi entropy has been proposed. We study their practical applicability to engineering secure quantum communication systems. By deriving a composable security bound using sandwiched R\'{e}nyi entropy, we apply our framework to a practical scenario involving BPSK modulation over a lossy bosonic channel, consistent with DVB-S2X standards. We highlight a critical trade-off between reliability and security, by showing that current coding rates may fall short of supporting positive secure coding rates under physical layer security considerations. This emphasizes the need for higher coding rates or additional frame lengths to balance reliability-security trade-offs effectively. Through this application, we demonstrate the utility of $\alpha$-order measures for advancing secure communication in quantum physical layer designs.
\end{abstract}

\maketitle

\section{Introduction}
Security at the physical layer of wireless communication systems has garnered significant attention from both academia and industry. While not fully standardized in 5G \cite{Sun2019}, it remains a topic of active research. For 6G, physical layer security is expected to play a pivotal role, driven by increasing demands for privacy and security in ultra-reliable, low-latency communication systems, making security a mandatory feature by design \cite{Mucchi2021}. 
In particular, many researchers focus on 
the Digital Video Broadcasting – Satellite – Second Generation Extension (DVB-S2X) standard \cite{dvbs2x} as the new standard.
On the other hand, the theory of physical layer security
in the quantum regime has been well developed.
The aim of this perspective is 
to fill the gap between the theoretical results and 
practical scenarios based on DVB-S2X.

The theory of physical layer security has a long history as follows.
The classical approach to physical layer security, rooted in Wyner's wiretap channel model \cite{Wyner1975,Csiszar1978}, prioritizes secrecy capacity as the primary metric. However, the cryptographic community has demonstrated that secrecy capacity alone fails to capture operational guarantees, particularly in non-asymptotic settings \cite{Maurer1993}. 
For quantum setting, 
the papers \cite{CWY,Devetak} introduced the concept of classical-quantum wiretap channel.
The references \cite{DS} \cite[Eq. (9.62)]{Hayashi06} introduced the concept of the degraded channel in the classical-quantum scenario.
Although 
the capacity of classical wiretap channel has single-letterized expression \cite{Csiszar1978},
the obtained capacity of classical wiretap channel does not have
single-letterized expression \cite{DS,MH15-1}
because the maximum of the difference of two mutual information
does not satisfy the additivity property \cite{LWZG,TBR}.

For its practical use, the following two key components are needed.
One is the finite-length security evaluation, and the other is the computationally efficient code construction.
To resolve these problems, 
the paper \cite{Hayashi2011} proposed the method by using 
universal2 hash function \cite{CW79,Krawczyk} for the classical wiretap channel.
For this aim, 
the paper \cite{Hayashi2011} extended the leftover hashing lemma \cite{Bennett1995,HILL} to the version with relative R\'{e}nyi entropy of a general order.
That is, the original leftover hashing lemma \cite{Bennett1995,HILL} employs only 
relative R\'{e}nyi entropy of order 2.
The extended leftover hashing lemma \cite{Hayashi2011} 
employs relative R\'{e}nyi entropy whose order runs from $2$ to $1$.
Then, the paper \cite{Hayashi2011}
constructed a code for wiretap channel
by combining linear error correcting code and 
a universal2 hash function \cite{CW79,Krawczyk}.
Although the paper \cite{Hayashi2011} employs relative entropy as a security measure, 
the paper \cite{H13} reformulate this result in a way to adopt
the universally composable security measure. 
The paper \cite{YT2019} also used the (normalized or unnormalized) 
R\'{e}nyi divergence to measure the level of approximation. 
The paper \cite{Endo2016,GLZOZ,FIKET,WD18,JHCS,Hain17-3,EFKT,Hayashi2020,HV20,HMWJ,SYYTY} applied this method for classical wiretap channel to free-space communication including satellite channels.

Toward its classical-quantum scenario, 
the paper \cite{RW05,Renner} established the classical-quantum version of 
The original leftover hashing lemma by \cite{Bennett1995,HILL}.
The paper \cite{H15} derived
classical-quantum leftover hashing lemma with general order 
under the relative entropy criterion by extending the result by 
\cite{Hayashi2011}.
Later, the paper \cite{MH14} tried to extend the result by \cite{H13} to 
the classical-quantum setting, but
the obtained upper bound is weaker than the classical case \cite{H13}.
The paper \cite{MH15-1} applied these results to 
the classical-quantum wiretap channel.
Recently, Dupuis \cite{Dupuis} succeeded in extending
the result by \cite{H13} to the classical-quantum setting
by using the sandwich version of quantum relative R\'{e}nyi entropy \cite{Wilde_2014,M_ller_Lennert_2013}.
Using the result \cite{Dupuis},
the paper \cite{WLH} established 
computationally efficient codes for  
classical-quantum wiretap channel 
whose security evaluation is given 
in finite-length regimes under the universally composable security measure. 
Recently, the papers \cite{PSCA,AngQtm21,AngQtm24}
applied the classical-quantum wiretap channel model
to realistic scenarios.

The aim of this perspective is to apply the above mentioned 
computationally efficient codes for classical-quantum wiretap channel 
to a practical and realistic engineering scenario based on DVB-S2X, which leads us to
the development of quantum physical layer security systems for real-world implementations.
To achieve this aim, this perspective covers the following two topics.
First, we provide a comprehensive review of the so-called $\alpha$-order information-theoretic metrics, highlighting their advantages over traditional secure information metrics, being particularly relevant that they offer composable security guarantees, ensuring robust and meaningful system-level security. 
Second, we demonstrate that, unlike traditional metrics such as secrecy capacity, which lack operational meaning in non-asymptotic settings, $\alpha$-order metrics enable precise quantification of the trade-offs between reliability and security, positioning them as practical tools for realistic system design (in the finite-length regime). 

To illustrate our claimed operational utility of the $\alpha$-order information-theoretic metrics, we present an example in the domain of space communications, where secure and reliable transmission is critical. Specifically, we apply our proposed $\alpha$-order metrics to a scenario involving Binary Phase Shift Keying (BPSK) modulation over a lossy bosonic channel, 
aligned with the Digital Video Broadcasting – Satellite – Second Generation Extension (DVB-S2X) standard. This application demonstrates the ability of $\alpha$-order metrics to quantify security guarantees under realistic constraints, offering a pathway to integrate advanced security measures into existing communication frameworks.

Through this work, 
in order to bridge the gap between quantum information theory and practical engineering applications,
advanced information metrics, such as $\alpha$-order metrics, are presented not merely as theoretical constructs but as operational tools for designing secure systems. 
Our proposed methods stress the fact that \textbf{$\alpha$-order metrics provide quantifiable and composable security guarantees}, paving the way for their adoption in next-generation communication systems where security is a fundamental requirement.

The contents can be summarized as follows:
\begin{itemize}
    \item \textbf{Section 2}: Introduces our secrecy system model and the general construction of the secrecy code. It also introduces the reliability and the secrecy metrics, the latter being the distinguishability of two quantum states possessing the universal security composability property.
    \item \textbf{Section 3}: Reviews $\alpha$-order information-theoretic metrics and explains their advantages over traditional metrics, including composability, tighter bounds, and their ability to balance security and reliability in finite-length scenarios.
    \item \textbf{Section 4}: Provides the practical application of $\alpha$-order metrics to the design of privacy amplification in quantum physical layer security systems when the task is to extract secure keys.
    \item \textbf{Section 5}: Extends the practical application of $\alpha$-order metrics to the design of privacy amplification for our secrecy system model, i.e. when the task is to extract a secure message. This is accomplihsed by balancing reliability and security with the practical construction of the secrecy code to the constraints of the given coding rates and channel conditions.
    \item \textbf{Section 6}: Provides numerical results demonstrating how $\alpha$-order metrics quantify the reliability-security trade-offs in a DVB-S2X scenario assuming BPSK modulation. These results highlight the need for higher coding rates to achieve positive secure coding rates under stringent physical layer security requirements.
    \item \textbf{Section 7}: Concludes by summarizing key findings and outlining directions for future work to optimize the application of $\alpha$-order metrics for secure communication in quantum physical layer designs.
\end{itemize}

\section{Secrecy System Model}
\label{Sec:SystemModel}
\begin{figure*}
\centering
\includegraphics[width=\textwidth]{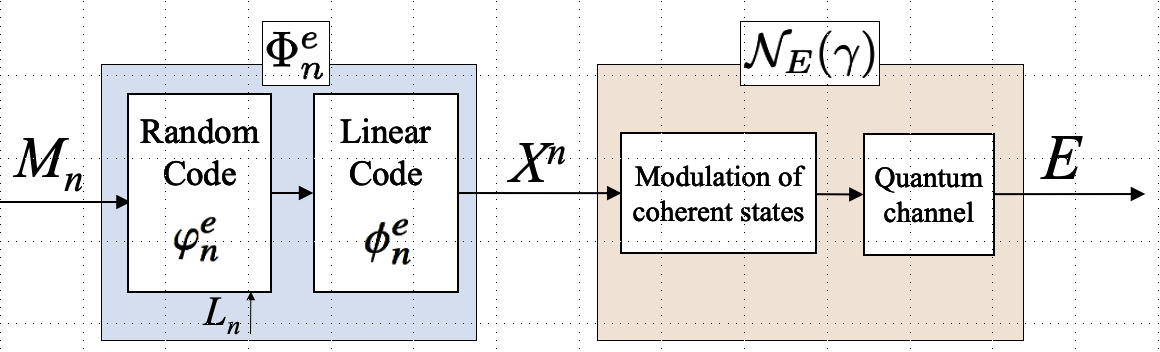}
\caption{Quantum Wiretap Secrecy System Model}
\label{fig:SystemModel}
\end{figure*}

Differently from the classical wiretap system originally proposed by Wyner and Csisz\'ar and K\"orner in \cite{Wyner1975}\cite{Csiszar1978}, 
we build our system model as
semi-quantum wiretap channel model that has two output systems,
the legitimate receiver's system and the eavesdropper's system.
In our model, 
the legitimate receiver's system is a classical system
because we fix the legitimate receiver's detector in our model.
But, to cover a powerful eavesdropper, 
the eavesdropper's system is modeled as a quantum system ${\cal H}_Z$ as Fig. \ref{fig:SystemModel}. 

A legitimate transmitter has a set of secret messages to be sent to the legitimate receiver, denoted as ${\cal M}=\{1, 2, \ldots , M\}$. 
%The single random variables of the legitimate transmitter and receiver are denoted as $X$ and $Y$, respectively, while the random variable at the eavesdropper receiver is denoted as $Z$. 

%The corresponding codewords are denoted as $X^n$, $Y^n$, $Z^n$ from sets $\mathcal{X}$, $\mathcal{Y}$, and $\mathcal{Z}$, respectively. 

The random variable $L\in {\cal L}=\{1, 2, \ldots , L \}$ induces a randomized output $X^n$ independent of $M\in {\cal M}$. 
The modulated codewords are subject to equal power constraints for each codeword, with per-symbol power designed to comply with the link budgets. 
We denote the information message input probability as $p_M$.
The system assumes a broadcast channel model described as $W^{n}_{YZ|X}$. 
For $n$ uses of the channel, the reliability of the channel code and the secrecy of the random code are quantified as $\epsilon^B_n$ and $\delta^E_n$, respectively, and will be defined later below. 

The stochastic code makes use of known linear error-correcting codes and randomized hashing codes. The resulting composite stochastic code is denoted as\footnote{Randomization of the encoding map $\phi^e_n$ is driven by the random variable $L_n$. However, we do not make this random variable explicit in our notation.} 
$\Phi_n(\epsilon^B_n, \delta^E_n)$. 
We define a general construction as follows (an example of practical construction is presented in \cite{Hain17-3}).  
\begin{mydef}
\textbf{General construction of $\Phi_n(\epsilon_n, \delta_n)$}. A general construction of the composite code, $\Phi_n$, is given as the pair of encoding and decoding maps $(\Phi^e_n,\Phi^d_n)$ with $ \Phi^e_n : M_n \rightarrow \Phi^e_n(M_n)=X^n$ with $\Phi^e_n = \phi^e_n \circ \varphi^e_n $, where $\varphi_n^e$ denotes a random code providing secrecy quantified as $\delta_n$ and $\phi^e_n$ denotes a conventional deterministic linear code providing reliability quantified as $\epsilon_n$. The corresponding decoding map is $ \Phi^d_n : Y^n \rightarrow \Phi^d_n(Y^n)\in \cal M$ with  $\Phi^d_n = \varphi^d_n \circ \phi^d_n$.  
\end{mydef}
For the random encoding/decoding maps, we consider randomized hash functions,  $\mathbf{f_{L}}$, which are stochastic maps from a set $\cal{H}$ to $\mathcal{M} := \{1, 2, \dots M \}$, where $L$ denotes the variable for random selection. 

For each use of the quantum channel, the legitimate information transmitter (Alice) prepares a coherent state modulated by the random variable $X$.
%, where $X=0$ and $X=1$ happen with probability $q$ and $1-q$, respectively. 
After $n$ transmissions, Bob and Eve detect their arriving quantum signals. 
We assume that Bob may only have access to state-of-the-art (not perfect) detectors on each transmission
so that Bob's receiving system is formulated as the $n$-fold system of a classical system.
But, Eve is assumed to apply the best quantum detection strategy so that 
Eve's system is treated as the $n$-fold system of a quantum system.
%$Z^n$ (i.e., Helstrom detection for binary discrete signals). 
%Bob and Eve send their estimated received states for decoding since  $\Phi_n$ is publicly known. 
The stochastic code must be well-designed to guarantee that Bob's decoder satisfies the average error probability constraint
\begin{align}
\bar{\epsilon} (\Phi_n) = \bar{P}_e (\phi^d_n(Y^n)  \neq \phi^e_n(X^n) ) \leq \epsilon_n,
\end{align}
while also ensuring that Eve's cryptoanalysis of the information-transforming flow through the secrecy system model is sufficiently limited. 
Here, we assume a bipartite quantum state $\rho_{ME}$, 
with the message $M$ held by Alice and $E$ by the adversary Eve. 
The amount of information leakage is evaluated as 
%The random operation from a given hashing family $h$, which is publicly known, $\{\mathcal{R}_h^{M \to M'} : h \in \mathcal{H}\}$, is given by
\begin{align} \label{eq:delta_expression}
\bar{d}_h (\Phi_n) =  \left\| \rho_{ME} - \rho_M \otimes \rho_E \right\|_1 \leq \delta_n.
\end{align}
Here, $M$ is a classical system and $E$ is a quantum system so that
$\rho_{ME}$ is treated as a classical-quantum state.
%where $M$ represents the message system to be transmitted, $E$ represents Eve's quantum system when the message $M$ is sent using the code $\Phi_n(\epsilon_n, \delta_n)$, and $\rho_{ME}$ represents the joint state. 
This metric quantifies the distinguishability of two quantum states and also possesses the universal composability property \cite{Renner}.
When this random variable is used as a component of a cryptographic protocol, it indicates the deviation from the case of a perfectly secure random number in terms of information-theoretical security.

The security metric \eqref{eq:delta_expression} can be upper bounded using various quantum information-theoretical metrics, as discussed in the next section. In particular, we focus on metrics based on $\alpha$-order entropies, which offer several advantages over traditional smoothing and min-entropy approaches. These $\alpha$-order metrics are not only more straightforward to apply but also provide tighter security bounds in practical scenarios. Their simplicity and effectiveness in capturing the trade-offs between secrecy and reliability make them a superior choice for engineering design.

We say that $\Phi_n(\epsilon_n, \delta_n)$ is $(\epsilon_n, \delta_n)$-achievable if $\lim_{n \rightarrow \infty} \bar{\epsilon} (\Phi_n) = 0$ and $\lim_{n \rightarrow \infty}  \bar{d}_h (\Phi_n) = 0$. 
For the security metric, we usually also consider the exponential decreasing rate (exponent) defined as
\begin{align}
e_d(\Phi_n) =  -\frac{1}{n} \log \bar{d}_h (\Phi_n).
\end{align}

\section{$\alpha$-order Quantum Information Theoretic Metrics}
A fundamental challenge in designing secure communication systems is quantifying the trade-offs between reliability and security, particularly in non-asymptotic scenarios. Traditional metrics, such as smoothing and min-entropy, often fail to capture these trade-offs effectively under practical constraints. To address this gap, $\alpha$-order information-theoretic metrics based on Rényi entropy provide a unified framework for analyzing security and reliability in finite-length regimes, making them well-suited for both theoretical and practical applications.

In this section, we present a detailed review of $\alpha$-order metrics and their relevance to quantum physical layer security. Specifically, we highlight how these metrics generalize classical measures, such as mutual information and relative entropy, to the quantum domain. The discussion also includes an exploration of their operational advantages, such as tighter security bounds, robust performance under finite-length constraints, and alignment with composable security principles.

\subsection{Definition of information-theoretic security}
In our work, we presume that Eve is a passive eavesdropper, which means that she cannot interfere with the channel actively by inserting or modifying messages undetected. This presumption can be justified by using well-known authentication techniques that are unconditionally secure, provided that Alice and Bob share a short, unconditionally secure authentication method.
\begin{mydef}
\textbf{Information-theoretical security}. A protocol provides this type of security when it is secure against any adversary, regardless of their computational resources. It is achieved by ensuring that Eve has zero or negligibly small information about the plaintext, based on the transmissions Eve can intercept and measure.
\end{mydef}
The primary characteristic of information-theoretical security is that the security does not rely on unproven assumptions about computational problems, and therefore it can be regarded as quantum-safe with respect to potential quantum computing threats. This is in contrast to \textbf{computational security}, which would claim that e.g. the hash functions used in our secrecy system model are quantum-safe because while quantum computing reduces the effective security level of hash functions by roughly a factor of two (turning a 256-bit hash into the security level of a 128-bit one), it does not completely break them like it potentially does with e.g. current asymmetric cryptographic algorithms. 

%Hence, we want to provide information-theoretical security guarantees to the physical communication process of our secrecy system model. 

Hence, we need to identify quantum information theoretical security metrics and corresponding security criteria that it allow us to guarantee that Eve will not be able to correctly obtain the message transmitted by Alice, for any meaningful (operational) claim to security. 

Next, we review different Rényi information theoretical security measures relevant to quantifying security.

\subsection{Quantum Relative Entropy}
Quantum relative entropy for a state $\rho$ and a positive semi-definite operator $\sigma$ is defined as:
\[
D(\rho \| \sigma) := \text{Tr}[\rho(\log_2 \rho - \log_2 \sigma)],
\]
which generalizes the concept of classical relative entropy to the quantum domain. This metric plays a fundamental role in various quantum information processing tasks, such as distinguishing quantum states and hypothesis testing.

\subsection{Petz Rényi Divergence}
The Petz Rényi Divergence (or relative entropy), a generalization of classical Rényi relative entropy to the quantum case, is defined for $\alpha \in (0, 1) \cup (1, \infty)$ as:
\[
D_\alpha(\rho \| \sigma) := \frac{1}{\alpha - 1} \log_2 \text{Tr}[\rho^\alpha \sigma^{1-\alpha}].
\]
This measure converges to the quantum relative entropy as $\alpha \rightarrow 1$ and satisfies a data-processing inequality for $\alpha \in (0, 1) \cup (1, 2]$. However, this metric does not have convenient properties such as limited range for data-processing inequality, non-convexity in its second argument 
 for general values of $\alpha$, can be sensitive to small variations in the states and it does not always align well with operational tasks such as quantum hypothesis testing and quantum channel discrimination.

\subsection{Sandwiched Rényi Divergence}
The Sandwiched Rényi Divergence \cite{Wilde_2014,M_ller_Lennert_2013} was developed to address the limitations of the Petz–Rényi divergence and has several good operational properties. It is defined for $\alpha \in (0, 1) \cup (1, \infty)$ as:
\[
\tilde{D}_\alpha(\rho \| \sigma) := \frac{1}{\alpha - 1} \log_2 \text{Tr}\left[\left(\sigma^{\frac{1-\alpha}{2\alpha}} \rho \sigma^{\frac{1-\alpha}{2\alpha}}\right)^\alpha\right].
\]
This metric also converges to the quantum relative entropy as $\alpha \rightarrow 1$, and it satisfies a data-processing inequality for all $\alpha \in [\frac{1}{2}, 1) \cup (1, \infty)$. Both the Petz–Rényi and sandwiched Rényi relative entropies possess important properties such as additivity and ordering (i.e., for $\alpha > \beta > 0$, $D_\alpha(\rho \| \sigma) \geq D_\beta(\rho \| \sigma)$ and $\tilde{D}_\alpha(\rho \| \sigma) \geq \tilde{D}_\beta(\rho \| \sigma)$). 

The Sandwiched Rényi Divergence finds wider applicability due to better adherence to the Data Processing Inequality (i.e. it decreases under quantum operations), stronger operational connection (especially in hypothesis testing and channel discrimination), convexity in its second argument, aiding in optimization problems. It shows also a smoother behavior and better alignment with quantum relative entropy. Finally, it provides a well-defined basis for conditional Rényi entropy and mutual information, which are useful in evaluating secrecy and security in quantum information tasks as we aim in this work and we present next.

\subsection{The $\alpha$-order Mutual Information}
In this case, we employ the sandwiched Rényi divergence to define the $\alpha$-order mutual information $I_\alpha(A;E)_\rho$ of a bipartite quantum state $\rho_{AE}$ as:
\[
I_\alpha(A;E)[\rho_{AE}] 
:= \inf_{\sigma_E \in \mathcal{D}(H_E)} \tilde{D}_\alpha(\rho_{AE} \| \rho_A \otimes \sigma_E),
\]
where $\mathcal{D}(H_E)$ denotes the set of density operators on the Hilbert space $H_E$ and we consider the interval $1 \leq \alpha \leq 2$ for the following reason. For our work, there is no need to extend the bound to $\alpha > 2$ because the randomization induced in the stochastic communication flow is only of the second moment of the state. The $\alpha$-order mutual information $I_\alpha(A;E)[\rho_{AE}]$ possesses several important properties such as monotonicity, data-processing inequality, additivity and for $\alpha \to 1$, converges to the standard quantum mutual information:
    \[
    \lim_{\alpha \to 1} I_\alpha(A;E)[\rho_{AE}] = I(A;E)[\rho_{AE}],
    \]
where $I(A;E)_\rho = D(\rho_{AE} \| \rho_A \otimes \rho_E)$ is the quantum mutual information. The $\alpha$-order mutual information has important operational meanings in quantum information theory. In quantum cryptography, its meaning is tied to the fact that the physical channels (due to environmental interactions, imperfections, or deliberate noise processes) randomize the $\alpha$-order moment of the state, i.e. how the channel interacts with the statistical structure of the state.

\section{Privacy amplification with $\alpha$-order conditional R\'{e}nyi entropy}
Building on the review of the theoretical foundation of $\alpha$-order metrics introduced in Section 3, we now explore their operational application to the design of privacy amplification in quantum physical layer security systems. In quantum cryptography, privacy amplification is a crucial step to ensure the secrecy of the transmitted information by reducing Eve’s knowledge about the secret message. To achieve such reduction we use the distinguishability metric that also possesses \textbf{the universal composability property} \cite{Renner} given as (2), which we reproduce here for convenience
\begin{align} \label{eq:delta_expression}
\left\| \rho_{ME} - \rho_{M} \otimes \rho_E \right\|_1 ,
\end{align}
where $M$ is the information of our interest
and $\rho_{ME}$ is the joint state between $M$ and the eavesdropper's information.

The most typical method to design privacy amplification is the application of universal 2 hash function. A randomized function $F$ from ${\cal A}$ to ${\cal M}$ is called 
a universal 2 hash function
when the relation
\begin{align}
{\rm Pr} (F(a)=F(a') ) \le \frac{|{\cal M}|}{|{\cal A}|}
\end{align}
for any two elements $a\neq a' \in {\cal A}$.

To handle the quantity \eqref{eq:delta_expression} with the application of 
a universal 2 hash function,
we focus on the conditional R\'{e}nyi entropy of order $\alpha$:
\begin{align}
H_{\alpha}(A|E)_\rho:= \max_{\sigma_E}\frac{1}{1-\alpha}\log \Tr{
\sigma_E^{\frac{1-\alpha}{2\alpha}}\rho_{AE}\sigma_E^{\frac{1-\alpha}{2\alpha}}}^\alpha.
\end{align}
Recently, using the completely mixed state $\rho_{M,mix}$ on the classical system ${\cal M}$,
given a classical-quantum state $\rho_{AE}$,
the paper \cite{Dupuis} showed the relation
\begin{align}
\mathbb{E}_F \left\| \rho_{F(A) E} - \rho_{M,mix} \otimes \rho_E \right\|_1% \notag\\
\le  2^{\frac{2}{\alpha}-1+ \frac{\alpha-1}{\alpha}
(\log |{\cal M}|-H_{\alpha}(A|E)_\rho)} \label{NMG}
\end{align}
for $\alpha>1$,
which is a quantum extension of \cite[Eq. (67)]{H13}.
The reference \cite[Eq. (67)]{H13} derived a similar relation 
by replacing the trace norm by quantum relative entropy in \eqref{NMG}
in the classical case, and 
the reference \cite{Dupuis} extended it to the quantum case.
Although the reference \cite[Eq. (91)]{MH14} considered its quantum extension before \cite{Dupuis},
the evaluation by \cite[Eq. (91)]{MH14} is worse than \cite{Dupuis}.
The case $\alpha=2$ was obtained by \cite{Renner}.

\section{Application to the Quantum Wiretap Secrecy System Model}
%However, the simple application of the above evaluation works only when we extract secure keys
The previous evaluation works only when the task is to extract secure keys from the random variable $A$, which is a different task to extract a secure message from the random variable $A$ in our quantum wiretap secrecy system model.
%this which is different problem from the wiretap channel.
As stated in \cite[Section V]{WLH}, in order to apply the above evaluation to the wiretap scenario, we need to apply the methods developed in the related literature, specifically in \cite[Section V]{Hayashi2011}, \cite[Section VIII]{H13} and \cite[Appendix A-B]{Hain17-3}. For the sake of completion, we summarize in the following the logical flow of derivations.

We denote the channel to Eve by $W_{Z|X}$, where $W_{Z|X=x}$ expresses the density matrix on Eve's system with Alice's input is $X=x\in \mathbb{F}_2$.
When Alice's input is given as $x^n=(x_1, \ldots, x_n)\in \mathbb{F}_2^n$, 
Eve's state is given as
\begin{align}
W^n_{Z|X=x^n}=W_{Z|X=x_1}\otimes \cdots \otimes W_{Z|X=x_n}.
\end{align}

Now, we consider the following scenario for a simple discussion.
For correct message transmission to Bob,
Alice chooses a linear subset ${\cal C} \subset 
\mathbb{F}_2^n$ with dimension $k_1+k_2$,
where $\phi^e_n$ denotes the linear encoding map
$\mathbb{F}_2^{k_1+k_2}$ to ${\cal C}$.
Also, we denote its decoder by $\phi^d_n$.
Then, Alice sends one element $A$ of ${\cal C}$.
In the following discussion, we assume that Bob decodes the transmitted message with the decoding error probability 
$\epsilon_n$.

Hence, we discuss only the secrecy to Eve.
The joint state across Alice's information and Eve's information is given as
\begin{align}
\rho_{AE}:= 2^{-(k_1+k_2)}\sum_{a \in {\cal C}}
|a\rangle \langle a| \otimes W^n_{Z|X=a}.
\end{align}
Then, we apply universal2 hash function $F$ from
${\cal C}$ to ${\cal M}:= \mathbb{F}_2^{k_1}$.
Denoting the information $F(A)$ by $M$, we have
\begin{align} 
\mathbb{E}_F\left\| \rho_{ME} - \rho_{M} \otimes \rho_E \right\|_1 
\le 2^{\frac{2}{\alpha}-1+ \frac{\alpha-1}{\alpha}
(k_1-H_{\alpha}(A|E)_{\rho_{AE}})},\label{NMG2}
\end{align}
where $H_{\alpha}(A|E)_{\rho_{AE}}$ is characterized as
\begin{align} 
&2^{-(\alpha-1) H_{\alpha}(A|E)_\rho}
=\min_{\sigma_E} \Tr{
\sigma_E^{\frac{1-\alpha}{2\alpha}}\rho_{AE}
\sigma_E^{\frac{1-\alpha}{2\alpha}}}^\alpha 
\end{align}
for $\alpha>1$.
By using $\rho_{{\cal C}}=2^{-(k_1+k_2)}\sum_{a \in {\cal C}}
|a\rangle \langle a|$,
this quantity is calculated as 
\begin{align} 
&2^{-(\alpha-1) H_{\alpha}(A|E)_\rho}\notag\\
=&
\min_{\sigma_E} \Tr
\Big(
\sigma_E^{\frac{1-\alpha}{2\alpha}}
\Big( 2^{-(k_1+k_2)}\sum_{a \in {\cal C}}
|a\rangle \langle a| \otimes W^n_{Z|X=a}\Big)
\sigma_E^{\frac{1-\alpha}{2\alpha}} 
\Big)^\alpha \notag\\
=&
2^{-(k_1+k_2)(\alpha-1)}
\min_{\sigma_E} \Tr
\Big(
(\rho_{{\cal C}}\otimes \sigma_E)^{\frac{1-\alpha}{2\alpha}}
\notag\\
&\cdot\Big( 2^{-(k_1+k_2)}\sum_{a \in {\cal C}}
|a\rangle \langle a| \otimes W^n_{Z|X=a}\Big)
(\rho_{{\cal C}}\otimes\sigma_E)^{\frac{1-\alpha}{2\alpha}} 
\Big)^\alpha .
\end{align}
Thus, we have
\begin{align} 
& H_{\alpha}(A|E)_\rho \notag\\
=&
k_1+k_2-
\min_{\sigma_E}\tilde{D}_\alpha (2^{-(k_1+k_2)}\sum_{a \in {\cal C}} |a\rangle \langle a| \otimes W^n_{Z|X=a}
\|\rho_{{\cal C}}\otimes\sigma_E) \notag\\
=&
k_1+k_2-
I_\alpha(A;E)[2^{-(k_1+k_2)}\sum_{a \in {\cal C}} |a\rangle \langle a| \otimes W^n_{Z|X=a}] \notag\\
\ge &
k_1+k_2-
\max_{Q_n}
I_\alpha(X;E)\Big[
\sum_{x^n \in \mathbb{F}_2^n} 
Q_n(x^n)|x^n\rangle \langle x^n| \otimes W^n_{Z|X=x^n}\Big] \notag\\
\stackrel{(a)}{=}&
k_1+k_2-
n \max_{Q}
I_\alpha(X;E)\Big[
\sum_{x \in \mathbb{F}_2} 
Q(x)|x\rangle \langle x| \otimes W_{Z|X=x}\Big] ,\label{NMG4}
\end{align}
where $(a)$ follows from Lemma 7 of \cite{WLH}.
In the following, we simplify
$I_\alpha(X;E)\Big[
\sum_{x \in \mathbb{F}_2} 
Q(x)|x\rangle \langle x| \otimes W_{Z|X=x}\Big] $ to 
$I_\alpha(X;E)_Q $.
Combining \eqref{NMG2} and \eqref{NMG4}, we have
\begin{align} 
\mathbb{E}_F\left\| \rho_{ME} - \rho_{M} \otimes \rho_E \right\|_1 %\notag\\
\le 2^{\frac{2}{\alpha}-1+ \frac{\alpha-1}{\alpha}
(n\max_Q I_\alpha(X;E)_Q-k_2)}. \label{NMG3}
\end{align}

However, the above procedure cannot be considered as 
a code for wiretap channel
because Alice makes privacy amplification after her message transmission.
Hence, we need to modify the above protocol to generate a wire-tap code.

We choose a 
$k_1\times k_2$ Toeplitz matrxi $T(S)$
with random seed $S$. 
Then, using the idea \cite[Appendix A-B]{Hain17-3},
we choose a random code
$\varphi_n^e$ from 
$\mathbb{F}_2^{k_1}$ to
$\mathbb{F}_2^{k_1+k_2}$ as follows.
Given an element $M \in \mathbb{F}_2^{k_1}$,
we randomly choose an element $L\in \mathbb{F}_2^{k_2}$
and output 
$\left(
\begin{array}{cc}
I & -T(S) \\
0& I
\end{array}
\right)
\left(
\begin{array}{c}
M \\
L
\end{array}
\right)
\in \mathbb{F}_2^{k_1+k_2}$.
For Bob's side, we choose the map $\varphi^d_n$
from 
$\mathbb{F}_2^{k_1+k_2}$ to $\mathbb{F}_2^{k_1}$ as follows.
$\left(
\begin{array}{c}
M \\
L
\end{array}
\right)
\mapsto 
(I ~T(S)) \left(
\begin{array}{c}
M \\
L
\end{array}
\right)$, where
$L$ is subject to the uniform distribution.
Then,
we construct the encoder as
$\Phi^e_n = \phi^e_n \circ \varphi^e_n $
and the decoder as $\Phi^d_n = \varphi^d_n \circ \phi^d_n$,
the decoding error probability of 
 the pair of encoding and decoding maps $(\Phi^e_n,\Phi^d_n)$ 
is upper bounded by 
$\epsilon_n$
because the relation 
\begin{align}
(I ~T(S)) \left(
\begin{array}{cc}
I & -T(S) \\
0& I
\end{array}
\right)
\left(
\begin{array}{c}
M \\
L
\end{array}
\right)=M
\end{align}
guarantees that $\varphi^d_n \circ \varphi^e_n$ 
is the identity map. 

In fact, the secrecy of this code is evaluated by using 
\eqref{NMG3} as follows.
Since the subspace ${\cal C}$ is a $k_1+k_2$-dimensional space, ${\cal C}$ can be identified with $\mathbb{F}_2^{k_1+k_2}$ via the map $\phi^e_n$ and $(\phi^e_n)^{-1}$.
Via this identification, 
we denote 
$\left(
\begin{array}{cc}
I & -T(S) \\
0& I
\end{array}
\right)
\left(
\begin{array}{c}
M \\
L
\end{array}
\right)
\in \mathbb{F}_2^{k_1+k_2}$ by $A\in {\cal C}$.
Due to the construction, 
$A$ is subject to the uniform distribution ${\cal C}$.
Since the map $\varphi^d_n$ 
forms a universal 2 hash function $F$ from
${\cal C}$ to ${\cal M}$, 
\eqref{NMG3} can be applied to the above randomly chosen 
wire-tap code $(\Phi^e_n,\Phi^d_n)$.

%\section{Proposed $\alpha$-order composable security criterion}
%This metric is difficult to compute exactly and it is convenient to use upper bounds. 

In order to simplify our evaluation, let's first introduce the following definition \cite{Hayashi2020}.
\vspace{0.2cm}
\begin{mydef}
\textbf{Sacrifice rate, $ \rho_{{\rm sac}}  $}. Given a composite code $\Phi(\epsilon_n, \delta_n)$ and $n$ uses of the channel, we call the \emph{sacrifice coding rate}, $\rho_{{\rm sac}}$, the extra transmission rate that needs to be used to guarantee a given level of secrecy as quantified by $\delta_n$.
\end{mydef}
\vspace{0.2cm}
Hence, the \emph{secure coding rate}, $\rho_{{\rm sec}}$, of the composite stochastic code $\Phi(\epsilon_n, \delta_n)$ is limited by the sacrificed rate as follows
\begin{align}
\rho_{{\rm sec}}(\Phi_n) &= \rho_{{\rm rel}}(\epsilon_n) - \rho_{{\rm sac}}(\delta_n),
\label{eq:SecRate}
\end{align}
where $ \rho_{{\rm rel}}(\epsilon_n) $ is the rate of the linear channel code. 
Next, we aim to quantify the guaranteed security of our protocol as a function of the controllable design parameter in our protocol, $\rho_{{\rm sac}}(\delta_n)$, which means we can guarantee \textbf{security by design}.

%Our proposed upper bound \cite{WLH} makes use of 
%$\alpha$-order measures, and 
Hence, the evaluation of \eqref{NMG2} according to the bound \eqref{NMG3} can be expressed as
follows:
\begin{align}
\delta_{n} \leq \min_{1 \leq \alpha \leq 2} 2^{\frac{2-\alpha}{\alpha}} 2^{-\frac{\alpha-1}{\alpha} \cdot n \left(\rho_{\text{sac}} - \max_Q I_{\alpha}(X;E)_Q\right)}.
\end{align}
We then make the following interesting observation: while our information leakage metric is the trace distance, which quantifies the distinguishability between two quantum states and ensures universal composability, our bound establishes a connection between the desired exponentially vanishing distinguishability and the amount of $\alpha$-order mutual information that Eve can extract. Moreover, we can control by design that the wiretap code makes such information leakage exponentially vanishing.

More especifically, when the value
$\left(\rho_{\text{sac}} - \max_Q I_{\alpha}(X;E)_Q\right)$
 is positive,
 the above value goes to zero exponentially.
In particular, 
$\max_Q I_{\alpha}(X;E)_Q$ is monotonically increasing for $\alpha>1$,
the limiting case $\alpha\to 1$ gives the best evaluation for 
 the sacrifice rate under the use of the above inequality.
That is, this case shows that 
 the sacrifice rate $\rho_{{\rm sac}}=\max_Q I (X;E)_Q$
 realizes the asymptotic secrecy.
However, in the finite-length setting,
we need to handle the following trade-off.
When $\alpha$ is close to $1$, 
the term $\left(\rho_{\text{sac}} - \max_Q I_{\alpha}(X;E)_Q\right)$
becomes larger, but
the term $\frac{\alpha-1}{\alpha}$ becomes smaller. Since the term $2^{\frac{2-\alpha}{\alpha}}$ is negligible, our problem is the maximization of the exponent
\begin{align}
e_d(\Phi_n) = \max_{1\le \alpha \le 2}
 \frac{\alpha-1}{\alpha} \left(\rho_{\text{sac}} - \max_Q I_{\alpha}(X;E)_Q\right).
\end{align}

Since the maximization is realized by a value except 
for $\alpha=2$ in many cases,
the above maximization for $\alpha\in[1,2]$
improves the exponent over the case $\alpha=2$.
In addition, when we assume the symmetry, i.e., 
there exists a unitary $U$ such that
\begin{align}
U W_{Z|X=0} U^\dagger =W_{Z|X=1},
\label{NMI}
\end{align}
Theorem 2 of \cite{WLH} guarantees that 
the maximum $\max_Q I_{\alpha}(X;E)_Q$ is realized by the uniform distribution on $\mathbb{F}_2$.

\section{Numerical results for DVB-S2X}
\subsection{Preliminaries}
In this section we show how to obtain the above bound for the design of a secure DVB-S2X link (Digital Video Broadcasting - Satellite - Second Generation Extension) standard \cite{dvbs2x} according to our secrecy system model. 

\begin{figure*}
\centering
\includegraphics[width=\textwidth]{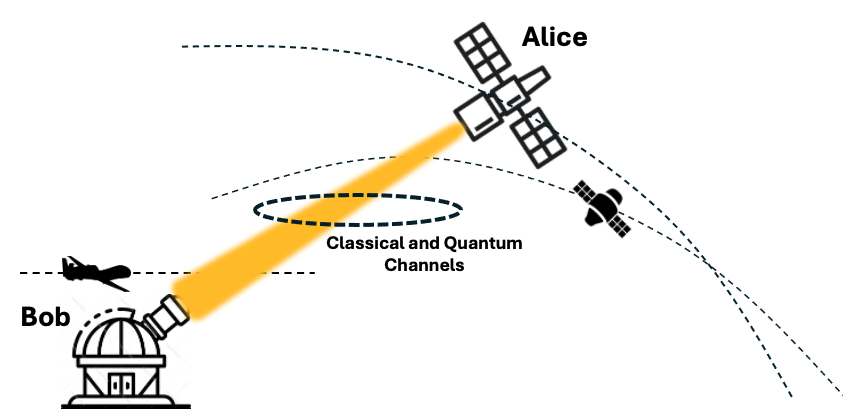}
\caption{Illustration of our space-based secure communication scenario showing the interaction between Alice (a satellite in Low Earth Orbit (ELO)) and Bob (a ground-based station) over a classical and quantum communication channel. The figure also exemplifies the challenges of secure space-based communication showing the presence of potential eavesdroppers (Eves), represented by a spy satellite and a small drone, which aim to intercept the communication link.} 
\label{fig:SpaceScenario}
\end{figure*}

DVB-S2X is an enhancement of the already established DVB-S2 standard, widely used for broadcasting media content via satellite. DVB-S2X facilitates greater flexibility and performance by introducing a variety of features, including better modulation and coding schemes, finer granularity in Forward Error Correction (FEC) rates, and additional roll-off options. These advancements allow for up to 20-30\% increase in efficiency compared to its predecessor, making DVB-S2X crucial for broadcasters aiming to optimize bandwidth and improve the quality of service, especially in the context of high-definition and ultra-high-definition broadcasts \cite{morello2016}. 

Our proposed security analysis is then useful for adding a security layer on top of current DVB-S2X guaranteed reliability, for example via multiplexing of classical and quantum signals. One of the key features of DVB-S2X is the definition of both short and long physical layer frames, which provide the flexibility required to adapt to different signal conditions and transmission needs. Short frames are beneficial in environments where low latency and fast decoding are essential. On the other hand, long frames are advantageous for achieving higher data throughput and robustness under weaker signal conditions.

Figure \ref{fig:SpaceScenario} illustrates the secure communication scenario considered in this work, where Alice, a satellite in Low Earth Orbit (LEO) which typically ranges from 160 km to 2000 km above the Earth's surface, communicates with Bob, a ground-based station, over classical and quantum channels, while contending with potential eavesdropping threats from spy satellites and drones that attempt to intercept the transmission. Note that a quantum link is more practical and efficient from a LEO satellite compared to higher orbits such as Medium Earth Orbit (MEO) or Geostationary Earth Orbit (GEO), as the shorter distance reduces channel loss and photon attenuation, enabling more reliable and secure quantum communication. However, it is unclear whether the DVB-S2X framing structure is optimal when multiplexing quantum and classical signals, this is a good topic for further research but out of the scope of this work. In the sequel, we do assume the current parameters in the standard. 

For the numerical evaluation, we focus on Very Short Frames consisting of $n = 16,200$ coded bits. Shorter frames are preferred here to ensure computational feasibility and precision in simulating the quantum wiretap channel. The coding rates available in the standard include:
\begin{equation*}
\rho_{\text{rel}}(\varepsilon_n^B) \in \left\{ \frac{2}{5}, \frac{1}{2}, \frac{3}{5}, \frac{2}{3}, \frac{3}{4}, \frac{4}{5}, \frac{5}{6}, \frac{8}{9}, \frac{9}{10} \right\}.
\end{equation*}
These rates offer a balance between robustness against errors (lower rates) and throughput (higher rates). For our operational use-case, we focus on the BPSK scenario, where the relevant rates are:
\begin{equation*}
\rho_{\text{rel}}(\varepsilon_n^B) \in \left\{ \frac{1}{2}, \frac{2}{5}, \frac{1}{3}, \frac{1}{4} \right\}.
\end{equation*}
BPSK is particularly advantageous in low SNR conditions, typical in satellite communications or environments with significant interference, making it a relevant choice for security analysis. Existing comparisons between DVB-S2X and 5G NTN systems, such as those summarized in \cite{5gntn1, 5gntn2}, suggest DVB-S2X's advantages in fixed scenarios, particularly for the downlink. However, further analysis is needed to evaluate its performance for LEO satellites under hybrid classical-quantum setups \cite{masini2022}.
% by Alice, namely binary phase shift keying (BPSK), producing coherent states $\ket{\alpha}_0$ and $\ket{\alpha}_1$. At Bob's detector arrive the coherent states $\ket{\sqrt{\eta}\alpha}$ and $\ket{-\sqrt{\eta}\alpha}$ and at Eve's detector arrive the coherent states $\ket{\sqrt{\gamma\eta}\alpha}$ and $\ket{-\sqrt{\gamma\eta}\alpha}$, where $\gamma$ measures the degradability of Eve's channel, denoted as $\mathcal{N}_E(\gamma),$ (see Fig. 1 of our secrecy system model) with respect to Bob's channel. BPSK is particularly useful for extremely robust, low-data-rate transmissions where the SNR is very poor, such as in certain satellite communication scenarios or in conditions with significant interference and thus a relevant security scenario as well.

\subsection{Security bound for BPSK}
Assuming equal probabilities of input states (i.e., the distribution of \(X\) is the uniform distribution on \(\{0, 1\}\)), it is convenient for the calculations to expand the BPSK quantum channel model
% in terms of orthogonal basis states as follows:
%\begin{align}
%x \in \{0, 1\} &\mapsto |v(x)\rangle:=\sqrt{p}|e_1\rangle + (-1)^x \sqrt{1-p}|e_2\rangle,
%\end{align}
% \(\{|e_1\rangle, |e_2\rangle\}\). 
%In such transformed domain, the parameter \(p\) characterizes the distribution of the energy of the BPSK states across the orthogonal states. 
with an orthogonal basis. Specifically, the received states of the BPSK modulation are given as:
\begin{align}
|\beta\rangle \text{ and } |-\beta\rangle \text{ with } \beta := \sqrt{\gamma\eta}\beta'.
\end{align}
whre $\beta'$ is the transmitted photonic power. We define the orthogonal basis as:
\begin{align}
|\varphi_e\rangle &:= (\cosh|\beta|^2)^{-1/2} \sum_{m=0}^{\infty} \frac{\beta^{2m}}{\sqrt{(2m)!}} |2m\rangle, \\
|\varphi_o\rangle &:= (\sinh|\beta|^2)^{-1/2} \sum_{m=0}^{\infty} \frac{\beta^{2m+1}}{\sqrt{(2m+1)!}} |2m+1\rangle.
\end{align}

Hence, using the relation \cite[eq. (12)]{HV}, we can express the received signals as:
\begin{align}
|\pm \beta\rangle &= \sqrt{\beta_e}|\varphi_e\rangle \pm \sqrt{\beta_o}|\varphi_o\rangle,
\end{align}
where
\begin{align}
\beta_e := e^{-|\beta|^2} \cosh|\beta|^2, \quad
\beta_o := e^{-|\beta|^2} \sinh|\beta|^2.
\end{align}
Note that in such transformed domain, the parameters $\beta_e$ and $\beta_o$ characterize the distribution of the energy of the BPSK states across the orthogonal states. 

As shown in Appendix \ref{S4C}, for $\alpha>1$, 
we can calculate the \(\alpha\)-order mutual information for BPSK 
%quantity $\max_Q I_{\alpha}(X;E)_Q$
as
%\begin{align}
%\max_Q I_{\alpha}(X;E)_Q
%&=I^{\text{BPSK}}_\alpha(X;E|p)\\
%&:= \frac{2\alpha - 1}{\alpha - 1} \log_2 \left( p^{\frac{\alpha}{2\alpha - 1}} + (1-p)^{\frac{\alpha}{2\alpha - 1}} \right)
%\label{NMD}.
%\end{align}
%This helps in quantifying the mutual information and understanding the security bounds. 
%With this transformation, 
%the \(\alpha\)-order mutual information \eqref{NMD} is rewritten
%as
\begin{align}
I^{\text{BPSK}}_\alpha(X;E|\beta) = \frac{2\alpha - 1}{\alpha - 1} \log_2 \left( \beta_e^{\frac{\alpha}{2\alpha - 1}} + \beta_o^{\frac{\alpha}{2\alpha - 1}} \right). \label{BBAA}
\end{align}
%This expression is calculated by substituting \(\beta_o\) into \(p\). 

\begin{figure*}[!ht]
    \centering
    \includegraphics[width=\textwidth]{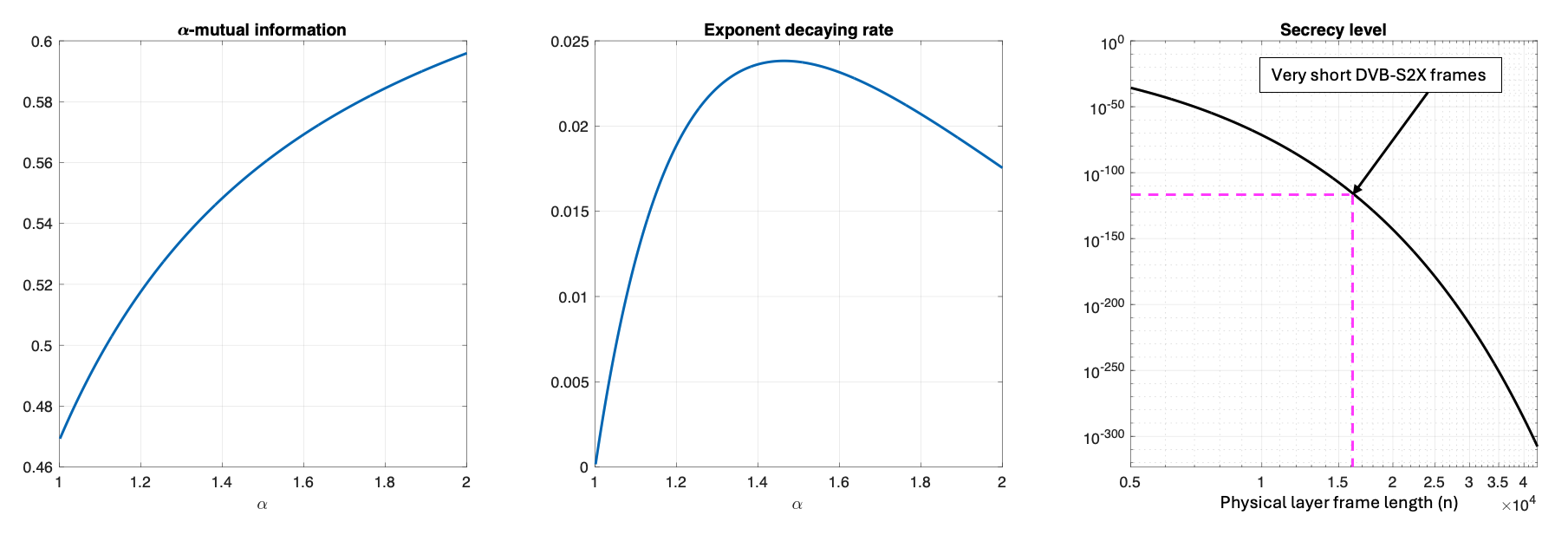}
\caption{Numerical values of our the (composable) information theoretical security metric. Left, values taken by the $\alpha$-mutual information. Centre, exponent numerical values. Right, secrecy level upper bound $\delta_{n}^{\text{BPSK}}$ for BPSK showing an example of operational point assuming DVB-S2X communication standard.}
\label{fig:Result_SPAWC}
\end{figure*}

If we denote $\delta_{n}^{\text{BPSK}}(\Phi_n |\alpha,\beta) $ the expression of the bound obtained in the previous section for the case of BPSK is
\begin{align}
e_d^{\text{BPSK}}(\Phi_n |\alpha,\beta) &=  -\frac{1}{n} \log\delta_{n}^{\text{BPSK}}(\Phi_n |\alpha,\beta) \notag\\
&=\frac{\alpha - 1}{\alpha} \left( \rho_{\text{sac}} - I^{\text{BPSK}}_\alpha(X;E|\beta)  \right).
\end{align}
Hence, the objective of the design is to optimize the value of \(\alpha\) that minimizes the value of the bound and \(\rho_{\text{sac}}\) to achieve a better (tighter) bound. 

\subsection{Numerical results}
In Fig. \ref{fig:Result_SPAWC} we present the numerical results assuming an average number of photons gathered by Eve $|\beta|^2 = 0.1$ and $\rho_{sac} = 0.631$. The first plot shows the values taken by the $\alpha$-mutual information. The second plot shows the exponent decaying rate. We observe the maximum rate occurring at $\alpha = 1.463865$  with a value of 0.0238. For this rate, the third plot of Fig. \ref{fig:Result_SPAWC}  shows that for Very Short Frames of DVB-S2X, the upper bound of the amount of leaked information is nearly zero. 

Hence, the rate $\rho_{{\rm sec}}(\Phi_n)$ is $(\epsilon^B_n, \delta^E_n)-$achievable for any $ \rho_{{\rm rel}}(\epsilon^B_n) >  \rho_{{\rm sac}}(\delta^E_n)$ with guaranteed quantifiable information theoretical security. Since the Normal Frame of DVB-S2X is much longer than the Very Short Frames, the rate $\rho_{{\rm sec}}(n, \Phi_n)$ is $(\epsilon^B_n, \delta^E_n)$ is also achievable for $ \rho_{{\rm rel}}(\epsilon^B_n) > \rho_{{\rm sac}}(\delta^E_n)$. 
%\begin{figure}[tbh]
%\centering
%\includegraphics[width=6cm]{images/ResultsTradeOff5}
%\caption{Secrecy-reliability trade-off for our DVB-S2 example with $n$ = 64800 coded bits and a $\rho_{{\rm rel}}(n,\epsilon^B_n)  %= 0.75$.}
%\label{fig:ResultsTradeOff}
%\end{figure}
\subsection{Reliability and security trafe-off for DVB-S2X}
It is important to note that (\ref{eq:SecRate}) reflects a trade-off between the guaranteed security against Eve and the guaranteed reliability for Bob. Specifically, we observe that

\begin{align}
\rho_{{\rm sec}}^{\text{BPSK}}(\Phi_n) &= \rho^{\text{BPSK}}_{{\rm rel}}(\epsilon_n) - 0.631 < 0,
\end{align}
due to the fact that $\rho^{\text{BPSK}}_{{\rm rel}}(\epsilon_n) \in \left\{ \frac{1}{2}, \frac{2}{5}, \frac{1}{3}, \frac{1}{4} \right\}$. To make DVB-S2X compatible with quantum physical layer security, higher coding rates, such as $\rho^{\text{BPSK}}_{{\rm rel}}(\epsilon_n) = 0.75$, would be necessary.

Our design and security analysis therefore highlights a critical trade-off between achieving physical layer security and maintaining the quality and robustness of the communication link, particularly when using BPSK in DVB-S2X.

However, the DVB-S2X standard includes specific coding rates that have been rigorously designed and tested to meet performance criteria, such as error rates and signal robustness, under typical operational conditions. The exclusion of a coding rate of $0.75$ from the standard likely stems from concerns that it might not offer sufficient error protection at the low SNRs commonly encountered with BPSK.

To address the need for a higher coding rate while maintaining security, potential solutions could include introducing additional frame lengths or modifying reliability targets. Shorter frames generally reduce latency and enable faster error correction, while longer frames are more efficient in terms of error correction and bandwidth usage. Adding new frame lengths could help find a compromise, but it would require careful consideration of the impact on overall system performance.

Alternatively, exploring more advanced error correction techniques that could allow for higher coding rates without compromising robustness might also be a viable solution. In any case, our recommendations aim to balance the desired security level with the performance metrics that are critical to the DVB-S2X standard.

\section{Conclusion and further work}
We have introduced a secrecy system model and developed a methodology for finite-length security analysis of quantum physical layer security protocols, proposing a composable security metric. Through a practical example involving a satellite communication standard, we demonstrated that our method enables the practical design of reliable and secure satellite communications over quantum states with a quantifiable, guaranteed level of secrecy. Our preliminary results highlight that while current satellite communication standards, which are primarily optimized for reliability, can integrate physical layer security, achieving this requires careful consideration of the trade-offs between security and reliability. Specifically, the inclusion of robust security guarantees may necessitate modifications to existing coding schemes or the adoption of new frame structures to maintain the desired balance between secure transmission and communication quality.

%%%%%
%%%%
%%%%

\appendix

\section{Derivation of \eqref{BBAA}}\label{S4C}
In order to prove \eqref{BBAA}, we employ the following 
parameterization of 
BPSK quantum channel model;
\begin{align}
x \in \{0, 1\} &\mapsto |v(x)\rangle:=\sqrt{p(1)}|e_1\rangle 
+ (-1)^x \sqrt{p(2)}|e_2\rangle,
\end{align}
with an orthogonal basis \(\{|e_1\rangle, |e_2\rangle\}\).
In this parameterzation, the relation \eqref{BBAA} is 
equivalent to the following relation;
\begin{align}
\max_Q I_\alpha(X;E)[\rho_{XE}^Q] 
=\frac{2\alpha-1}{\alpha - 1} \log_2 
\left(\sum_{z=1}^2 p(z)^{\frac{\alpha}{2\alpha-1}} \right)
.\label{NMA3}
\end{align}

For this derivation, we employ Reverse Holder inequality;
For $s>0$, we have
\begin{align}
\Big|\sum_{x} a(x) b(x) \Big| \ge 
\Big(\sum_x a(x)^{\frac{1}{1+s}}\Big)^{1+s}
\Big(\sum_x b(x)^{-\frac{1}{s}}\Big)^{-s}.
\end{align}
The quality holds when
$b(x)=C a(x)^{-\frac{1+s}{s}}$ with a constant $C$.
We consider distributions $Q'$ on $\{1,2\}$.
Reverse Holder inequality guarantees the following. 
Choosing $s=\frac{\alpha-1}{\alpha}$, we have
\begin{align}
&
\inf_{Q'}
\sum_{z=1}^2 p(z) Q'(z)^{\frac{1-\alpha}{\alpha}} 
=\inf_{Q'}
\Big(\sum_{z=1}^2 p(z)^{\frac{\alpha}{2\alpha-1}} 
\Big)^{\frac{2\alpha-1}{\alpha}} (\sum_{z=1}^2 Q'(z) )^{\frac{1-\alpha}{\alpha}} 
\notag\\
=&\Big(\sum_{z=1}^2 p(z)^{\frac{\alpha}{2\alpha-1}} 
\Big)^{\frac{2\alpha-1}{\alpha}}.\label{NCT}
\end{align}

Given a distribution, we have a state
$\rho_{XE}^Q :=\sum_{x \in \{0,1\}}
Q(x) |v(x)\rangle\langle v(x)| $.
For $\alpha>1$, we have
\begin{align}
&\max_Q I_\alpha(X;E)[\rho_{XE}^Q] 
= \max_Q \inf_{\sigma_E \in \mathcal{D}(H_E)} 
\tilde{D}_\alpha(\rho_{XE}^Q \| \rho_X \otimes \sigma_E) \notag\\
=&\max_Q \inf_{\sigma_E \in \mathcal{D}(H_E)} 
\frac{1}{\alpha - 1} \log_2 
\notag\\
&\Big(\text{Tr}
\sum_{x \in \{0,1\}}Q(x)
\left[\left(\sigma^{\frac{1-\alpha}{2\alpha}} |v(x)\rangle\langle v(x)| \sigma^{\frac{1-\alpha}{2\alpha}}\right)^\alpha\right] \Big)\notag\\
=&\frac{1}{\alpha - 1} \log_2 \text{Tr}
\max_Q \inf_{\sigma_E \in \mathcal{D}(H_E)} 
\sum_{x \in \{0,1\}}Q(x)\notag\\
&\cdot
\left[\left(\sigma^{\frac{1-\alpha}{2\alpha}} |v(x)\rangle\langle v(x)| \sigma^{\frac{1-\alpha}{2\alpha}}\right)^\alpha\right] \notag\\
=&\frac{1}{\alpha - 1} \log_2 \Bigg[
\max_Q \inf_{\sigma_E \in \mathcal{D}(H_E)} 
\sum_{x \in \{0,1\}}Q(x)
\notag\\
&\cdot\left(
\langle v(x)| \sigma^{\frac{1-\alpha}{\alpha}} |v(x)\rangle
\right)^\alpha \Bigg] \notag\\ %\label{NMA}
\stackrel{(a)}{=}&
\frac{1}{\alpha - 1} \log_2 \Bigg[
\frac{1}{2}
\inf_{\sigma_E \in \mathcal{D}(H_E)} 
\sum_{x \in \{0,1\}}
\left(
\langle v(x)| \sigma^{\frac{1-\alpha}{\alpha}} |v(x)\rangle
\right)^\alpha \Bigg] .\label{NMA}
\end{align}

The step $(a)$ can be shown as follows.
We choose the distribution $\tilde{Q}$ as $\tilde{Q}(x\oplus 1)=Q(x)$.
Since the symmetry for the exchange $x \to x\oplus 1$ implies 
\begin{align}
&\inf_{\sigma_E \in \mathcal{D}(H_E)} 
\sum_{x \in \{0,1\}}Q(x)
\left(
\langle v(x)| \sigma^{\frac{1-\alpha}{\alpha}} |v(x)\rangle
\right)^\alpha \notag\\
=& \inf_{\sigma_E \in \mathcal{D}(H_E)} \sum_{x \in \{0,1\}}\tilde{Q}(x)
\left(
\langle v(x)| \sigma^{\frac{1-\alpha}{\alpha}} |v(x)\rangle
\right)^\alpha,
\end{align}
we have
\begin{align}
&\inf_{\sigma_E \in \mathcal{D}(H_E)} 
\sum_{x \in \{0,1\}} \frac{Q(x)+\tilde{Q}(x)}{2}
\left(
\langle v(x)| \sigma^{\frac{1-\alpha}{\alpha}} |v(x)\rangle
\right)^\alpha \notag\\
\ge &
\frac{1}{2}\inf_{\sigma_E \in \mathcal{D}(H_E)} 
\sum_{x \in \{0,1\}}Q(x)
\left(
\langle v(x)| \sigma^{\frac{1-\alpha}{\alpha}} |v(x)\rangle
\right)^\alpha 
\notag\\
&+
\frac{1}{2}\inf_{\sigma_E \in \mathcal{D}(H_E)} 
\sum_{x \in \{0,1\}}\tilde{Q}(x)
\left(
\langle v(x)| \sigma^{\frac{1-\alpha}{\alpha}} |v(x)\rangle
\right)^\alpha \notag\\
=&\inf_{\sigma_E \in \mathcal{D}(H_E)} 
\sum_{x \in \{0,1\}}Q(x)
\left(
\langle v(x)| \sigma^{\frac{1-\alpha}{\alpha}} |v(x)\rangle
\right)^\alpha .
\end{align}
Since $ \frac{Q(x)+\tilde{Q}(x)}{2}=\frac{1}{2}$,
the maximum for $Q$ in \eqref{NMA}
is realized when $Q$ is realized when $Q(x)=1/2$.
\if0
\begin{align}
&\max_Q \inf_{\sigma_E \in \mathcal{D}(H_E)} 
\sum_{x \in \{0,1\}}Q(x)
\left(
\langle v(x)| \sigma^{\frac{1-\alpha}{\alpha}} |v(x)\rangle
\right)^\alpha
\notag\\
=&\frac{1}{2}
\inf_{\sigma_E \in \mathcal{D}(H_E)} 
\sum_{x \in \{0,1\}}
\left(
\langle v(x)| \sigma^{\frac{1-\alpha}{\alpha}} |v(x)\rangle
\right)^\alpha. \label{SY1}
\end{align}
\fi
Using the operator $V= |e_1\rangle \langle e_1|-|e_2\rangle \langle e_2|$, 
we have
\begin{align}
&\frac{1}{2}
\sum_{x \in \{0,1\}}
\left(
\langle v(x)| \sigma^{\frac{1-\alpha}{\alpha}} |v(x)\rangle
\right)^\alpha \notag\\
=&\frac{1}{2}
\sum_{x \in \{0,1\}}
\left(
\langle v(x)| V \sigma^{\frac{1-\alpha}{\alpha}} V |v(x)\rangle
\right)^\alpha \notag\\
=&\frac{1}{2}
\sum_{x \in \{0,1\}}
\frac{1}{2}\Bigg(
\left(
\langle v(x)|  \sigma^{\frac{1-\alpha}{\alpha}}  |v(x)\rangle
\right)^\alpha 
\notag\\
&+
\left(
\langle v(x)| V \sigma^{\frac{1-\alpha}{\alpha}} V |v(x)\rangle
\right)^\alpha
\Bigg) \notag\\
\ge &\frac{1}{2}
\sum_{x \in \{0,1\}}
\left(
\langle v(x)|  \frac{1}{2}
\sigma^{\frac{1-\alpha}{\alpha}} +\frac{1}{2}
V \sigma^{\frac{1-\alpha}{\alpha}} V |v(x)\rangle
\right)^\alpha \notag\\
\ge &\frac{1}{2}
\sum_{x \in \{0,1\}}
\left(
\langle v(x)|  
\left(\frac{1}{2}
\sigma +\frac{1}{2}
V \sigma^{\frac{1-\alpha}{\alpha}} V \right)^{\frac{1-\alpha}{\alpha}}
|v(x)\rangle
\right)^\alpha .\label{NMT8}
\end{align}
The state 
$\frac{1}{2}
\sigma +\frac{1}{2}
V \sigma^{\frac{1-\alpha}{\alpha}} V$
can be written as
$Q(1)|e_1\rangle \langle e_1|
+Q(2)|e_2\rangle \langle e_2|$
with a distribution $Q'$ on $\{1,2\}$.
Thus, we have
\begin{align}
&\frac{1}{2}
\inf_{\sigma_E \in \mathcal{D}(H_E)} 
\sum_{x \in \{0,1\}}
\left(
\langle v(x)| \sigma^{\frac{1-\alpha}{\alpha}} |v(x)\rangle
\right)^\alpha \notag\\
= &\frac{1}{2}
\inf_{Q'}
\sum_{x \in \{0,1\}}
\Bigg(
\langle v(x)|  
\Bigg(
Q'(1)|e_1\rangle \langle e_1|
\notag\\
&+Q'(2)|e_2\rangle \langle e_2|
\Bigg)^{\frac{1-\alpha}{\alpha}}
|v(x)\rangle
\Bigg)^\alpha \notag\\
= &\frac{1}{2}
\inf_{Q'}
\sum_{x \in \{0,1\}}
\Bigg(
\langle v(x)|  
\Bigg(
Q'(1)^{\frac{1-\alpha}{\alpha}}
|e_1\rangle \langle e_1|
\notag\\
&+Q'(2)^{\frac{1-\alpha}{\alpha}}|e_2\rangle \langle e_2|
\Bigg)
|v(x)\rangle
\Bigg)^\alpha \notag\\
= &
\inf_{Q'}
\left(
\sum_{z=1}^2 p(z) Q'(z)^{\frac{1-\alpha}{\alpha}} 
\right)^\alpha 
= 
\left(
\inf_{Q'}
\sum_{z=1}^2 p(z) Q'(z)^{\frac{1-\alpha}{\alpha}} 
\right)^\alpha \notag\\
\stackrel{(a)}{=}&
\left(
\left(\sum_{z=1}^2 p(z)^{\frac{\alpha}{2\alpha-1}} 
\right)^{\frac{2\alpha-1}{\alpha}}\right)^\alpha 
=\left(\sum_{z=1}^2 p(z)^{\frac{\alpha}{2\alpha-1}} \right)^{2\alpha-1},
\label{NMT}
\end{align}
where $(a)$ follows from \eqref{NCT}.

Therefore, combining \eqref{NMA} and \eqref{NMT},
we have
\begin{align}
&\max_Q I_\alpha(X;E)[\rho_{XE}^Q] \notag\\
=&\frac{1}{\alpha - 1} \log_2 \left[
\max_Q \inf_{\sigma_E \in \mathcal{D}(H_E)} 
\sum_{x \in \{0,1\}}Q(x)
\left(
\langle v(x)| \sigma^{\frac{1-\alpha}{\alpha}} |v(x)\rangle
\right)^\alpha \right] \notag\\
=&\frac{1}{\alpha - 1} \log_2 
\Bigg[\frac{1}{2}
\inf_{\sigma_E \in \mathcal{D}(H_E)} 
\sum_{x \in \{0,1\}}
\left(
\langle v(x)| \sigma^{\frac{1-\alpha}{\alpha}} |v(x)\rangle
\right)^\alpha \Bigg]\notag\\
=&
%\frac{1}{\alpha - 1} \log_2 
%\left(\sum_{z=1}^2 p(z)^{\frac{\alpha}{2\alpha-1}} \right)^{2\alpha-1}=
\frac{2\alpha-1}{\alpha - 1} \log_2 
\left(\sum_{z=1}^2 p(z)^{\frac{\alpha}{2\alpha-1}} \right),
\label{NMAU}
\end{align}
which implies \eqref{NMA3}.

% Acknowledgements
\medskip
\textbf{Acknowledgements} \par %delete if not applicable))
M.H. was supported in part by the National Natural Science Foundation of China (Grants no. 62171212). 
%Please insert your acknowledgements here

\medskip
\textbf{Conflict of Interest}
The authors declare no conflict of interest.

\medskip
\textbf{Data Availability Statement}
%Data sharing is not applicable to this article as no new data were created or analyzed in this study.
The code functions and simulation scripts used in this study are available from the corresponding author upon reasonable request.

% References
\medskip

% Use the following code if you wish to generate your bibliography with BibTeX;
% replace the string "MSP-template" below with the name(s) of
% the BibTeX data base(s) you want to use.
% The resulting bibliography-output (the content of the .bbl file)
% must be pasted back into this file before submission.
% Please also include your BibTeX data base file(s) in your submission
% so that we can re-run BibTeX if necessary.
%
%\bibliographystyle{MSP}
%\bibliography{MSP-template}

\end{document}